\documentclass{article}

\usepackage[final]{neurips_2022}

\usepackage[utf8]{inputenc} 
\usepackage[T1]{fontenc}    
\usepackage{hyperref}       
\usepackage{url}            
\usepackage{booktabs}       
\usepackage{amsfonts}       
\usepackage{nicefrac}       
\usepackage{microtype}      
\usepackage{xcolor}         

\title{Perspectives on the Social Impacts of \\
Reinforcement Learning with Human Feedback}

\author{%
  Gabrielle Kaili-May Liu\\
  Department of Mathematics \\
  Department of Brain and Cognitive Sciences\\
  Social and Ethical Responsibilities of Computing (SERC)\\
  Massachusetts Institute of Technology\\
  Cambridge, MA 02139 \\
  \texttt{gkml@mit.edu} \\
}

\begin{document}

\maketitle

\begin{abstract}
  Is it possible for machines to think like humans? And if it is, how should we go about teaching them to do so? As early as 1950, Alan Turing stated that we ought to teach machines in the way of teaching a child. Recently, reinforcement learning with human feedback (RLHF) has emerged as a strong candidate toward allowing agents to learn from human feedback in a naturalistic manner. RLHF is distinct from traditional reinforcement learning as it provides feedback from a human teacher in addition to a reward signal. It has been catapulted into public view by multiple high-profile AI applications, including OpenAI’s ChatGPT, DeepMind’s Sparrow, and Anthropic’s Claude. These highly capable chatbots are already overturning our understanding of how AI interacts with humanity. The wide applicability and burgeoning success of RLHF strongly motivate the need to evaluate its social impacts. In light of recent developments, this paper considers an important question: can RLHF be developed and used without negatively affecting human societies? Our objectives are threefold: to provide a systematic study of the social effects of RLHF; to identify key social and ethical issues of RLHF; and to discuss social impacts for stakeholders. Although text-based applications of RLHF have received much attention, it is crucial to consider when evaluating its social implications the diverse range of areas to which it may be deployed. We describe seven primary ways in which RLHF-based technologies will affect society by positively transforming human experiences with AI. This paper ultimately proposes that RLHF has potential to net positively impact areas of misinformation, AI value-alignment, bias, AI access, cross-cultural dialogue, industry, and workforce. As RLHF raises concerns that echo those of existing AI technologies for governance, industry, safety, ethics, and the future of global power relations, it will be important for all to be aware and intentional in the adoption of RLHF.
\end{abstract}

\section{Introduction}
Since long before modern computing we have sought to teach machines through natural, humanistic interactions \cite{GML5}. As early as 1950, Alan Turing stated in his seminal paper on artificial intelligence (AI) that we ought to ``provide the machine with the best sense organs that money can buy, and then teach it… That process could follow the normal teaching of a child. Things would be pointed out and named, etc.’’ \cite{TURING}. John McCarthy posed one of the earliest iteration of such a system in 1959, describing an ``advice taker’’ that could learn via common sense reasoning by drawing logical conclusions from any set of premises issued to the system as imperative statements \cite{MCCARTHY}. In the 1980s, this work was extended by Hayes-Roth et al. to develop a generalized framework for machines to learn from external (human) advice, involving steps for receiving, interpreting, and integrating advice into a machine’s learning \cite{HAYESROTH1, HAYESROTH2}. Since then, the rapid development of AI and machine learning (ML) has led to significant progress in giving artificial agents the ability to interact with humans and learn from their feedback in a naturalistic manner \cite{GML5, ME2}.

A technique of particular import which has arisen in the past few years is \textit{reinforcement learning with human feedback} (RLHF). Reinforcement learning (RL) refers to the field of ML in which an agent learns through interactions with the environment to select the best course of action (a policy) in a given state \cite{ME2, GML1}. Each state-action pair is described by a reward, which serves as feedback for the agent to tune its policy. As an agent learns through training episodes, it ultimately arrives at an optimized policy which permits maximization of the reward. RL has garnered high-profile success in various applications including board and video games, autonomous driving, text summarization, online personalization, finance, and healthcare. As such, it is thought to be a critical component in the development of truly generalized autonomous AI \cite{ME2}.

RLHF is an extension of RL that incorporates \textit{human feedback} into the training process \cite{GML1, OPENAI}. In addition to the reward signal, an RLHF agent receives feedback from a human teacher that permits it to learn with broader perspective and greater efficiency in a similar fashion to humans learning from the expertise of another human \cite{GML1}. By providing a bridge between an agent and a human teacher, RLHF allows humans to directly guide machine learning and machines to grasp elements of decision-making distinctly embedded in human experience \cite{RLHF8}. The ability to provide and incorporate human feedback in RLHF is further a critical step toward achieving improved alignment between ML models and human values \cite{GML1, RLHF1}.

Although RLHF is a relatively young technology, it has been catapulted into public view by multiple high- profile AI applications including OpenAI’s ChatGPT, DeepMind’s Sparrow, and Anthropic’s Claude. Uses of these chatbots include constructing context-appropriate email responses, solving math problems, and generating code \cite{RLHF6}. Presently, RLHF is finding widespread application in business, education, healthcare, and entertainment \cite{RLHF8}.

RLHF creates a host of benefits over traditional RL methods. Its key advantages lie in better alignment with human intentions, as well as planning conditional on future feedback, fluid learning from various types of feedback, and curation of feedback according to necessity, all of which are indispensable for creating truly intelligent agents \cite{GML1, RLHF2}. It also permits machines to learn by abstracting from what humans value as opposed to simply imitating human behavior, thereby equipping agents with greater adaptability, enhanced interpretability, and more reliable decision-making \cite{RLHF2}. 

Despite these advances, there is vast potential for RLHF to be improved \cite{RLHF1, FORBES}. RLHF models are potentially prone to inaccurate or harmful behavior (e.g., issuing racist statements) \cite{GML1}. This limitation reflects a longer-term challenge and motivation for improving RLHF \cite{OPENAI, RLHF1, FORBES}. Additionally, gathering human preference data as feedback is costly, and disagreement between human annotators adds variance to training data which can create confusion in situations in which the ground truth is obscure (e.g., ethical dilemmas) \cite{RLHF1}. Moreover, human feedback in RLHF is often constrained to be in the form of preference orderings which provide limited information and thereby restrict applicability \cite{GML1, RLHF2}. It is desirable to achieve a broader formalism that considers multiple types of feedback, dependent on task context and similar to the diversity of responses utilized in human learning. Work in this area could facilitate identification of which types of feedback lead to better generalization for applications. 

\subsection{Context for the Present Work}
As RLHF is gaining rapid traction, now is the ideal time to consider its potential impacts on society \cite{ME2}. The transformational potential of RLHF makes it critical to consider how broadened application of RLHF-based technologies may impact various stakeholders, what ethical concerns might arise as a result, how it may affect social and ethical challenges, and how governance may be utilized to mitigate risks. This analysis is timely and justified considering the current state of AI safety research. According to a 2022 report by the Center for Security and Emerging Technology, research in the top three areas of AI safety—robustness, interpretability, and reward learning—have seen explosive growth in the past decade \cite{CSET4}. Reward learning is critically concerned with reducing the risk of disparity between intended and observed outcomes, yet work in the area is less developed relative to robustness and interpretability research. This discrepancy extends to the study of related social and ethical implications \cite{CSET4}. This report therefore seeks to fill this gap and step toward expanding responsible discussion of reward learning methods like RLHF. 

\subsection{Objectives and Metrics}
The objective of this report is threefold: first, to provide a systematic study of the social effects of RLHF; second, to identify key social and ethical issues of RLHF; and third, to discuss social impacts for stakeholders. We propose that continued development of RLHF has a net positive social impact and is thus worth continued pursuit. We define an \textit{impact} to be any direct or indirect effect that ``includes both positive and negative results,’’ and a \textit{benefit} to be ``a positive impact that produces a good result’’ \cite{HELP1}. We assume \textit{social impact} to be the net effect on relevant stakeholders, which may include individuals, families, communities, organizations, nations, or global societies as a whole \cite{HELP1}. Social impacts identified in this report are evaluated relative to the \textit{social baseline}---the environment in the absence of the technology in question \cite{HELP1, SIA}.

\section{Impacts of RLHF}
We describe seven primary ways in which RLHF positively transforms human experiences with AI. Our analysis is guided by the following questions:
\begin{itemize}
\item How might RLHF affect the \textbf{integrity of information} to which people have access?
\item How might RLHF reflect \textbf{values and preferences} of target populations?
\item How might RLHF temper or intensify different axes of \textbf{social inequality}?
\item How might RLHF alter the \textbf{access} different social groups have \textbf{to AI technologies}?
\item How might RLHF impact \textbf{cultural and international relations}?
\item How might RLHF enhance \textbf{industries}?
\item How might RLHF transform \textbf{workforces and the organization of labor}?
\end{itemize}

\subsection{Combating Misinformation}
\begin{figure}[h!]
    \centering
    \includegraphics[width=0.75\textwidth]{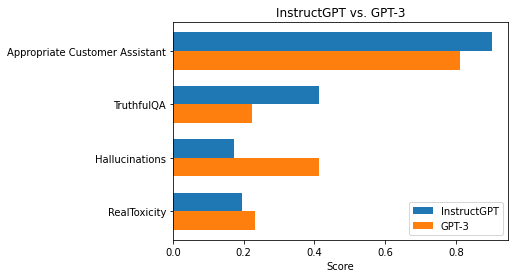}
    \caption{RLHF methods are significantly better versus state-of-the-art LLMs at mitigating toxic, false statements and generating truthful, appropriate content as indicated by the performance of InstructGPT versus GPT-3 \cite{GML1,OPENAI}.}
    \label{fig1}
\end{figure}

As an effective alignment technique, RLHF has significant potential to assist in mitigating harmful content generation that results from large language models (LLMs) and improve information integrity\footnote{\textit{Information integrity} refers to the dependability and trustworthiness of information \cite{FOOTNOTE}.}. LLM deficiencies are well-documented and range from biased outputs to leaked private data to misinformation and adversarial attack \cite{GML1}. Current approaches to moderating LLMs are cumbersome, require more data, or are overly complex \cite{GML1}. RLHF is a method that promises improved truthfulness and reduced toxicity of LLMs without significantly compromising performance or creating issues such as reduced representation of minorities in textual output \cite{GML1}. For instance, InstructGPT---trained with RLHF---exhibits enhanced ability versus GPT-3 to generate truthful and informative responses and follow unfamiliar instructions (Figure \ref{fig1}) \cite{GML1}. Combined with development of limitations on harmful output production, RLHF has great potential toward generation of positive content for assistive technologies, information sharing, and recommender/advice systems.

Even so, work remains to be done in order to improve the reliability of RLHF-based models. RLHF technologies like ChatGPT can still suffer from inappropriate and harmful outputs upon user request. The creators of ChatGPT and InstructGPT themselves described these technologies as perhaps being too obedient to user instruction \cite{GML1}. Such issues may be approached by combining RLHF with steerability methods or by modifying sampling procedures during training. It may also be useful to incorporate aspects of AI explainability, giving RLHF agents the ability to decline to comply with harmful requests and explain why they have done so \cite{GML2,morals}. Overall, what renders an output harmful often depends on context, and this can complicate model design \cite{GML1}.

Despite acting in broad alignment with human values, RLHF can still be misused for misinformation, oppression, or perpetuation of societal prejudice if guardrails are not established \cite{ME1}. RLHF-based content generation—whether visual, textual, auditory, or other forms of media—has application in disinformation and automated trolling, which can lead to compromised election integrity and public distrust in media and undermine the very fabric of organized governance in society. One can see that the power of RLHF in the disinformation context is threefold. First, human-machine teaming with RLHF models can accelerate message iteration, effectively increasing the productive power of disinformation campaigns \cite{CSET1}. Second, greater understanding of human behavior, preferences, and value systems could aid reconnaissance, leading to better mimicking of human activity, viewpoint manipulation, targeted messaging, conspiracy narrative generation, advancement of political narratives, and identification of feeble social fissures \cite{CSET5}. Third, knowledge that AI holds such potential may further erode trust and accelerate descent into the cynicism that advances disinformation \cite{CSET5}. 

\begin{quote}
\centering
\textit{``A people that no longer can believe anything cannot make up its mind. It is deprived not only of its capacity to act but also of its capacity to think and to judge. And with such a people you can then do what you please.’’} ---Hannah Arendt \cite{HA}
\end{quote}

There are a number of steps we can take to counter misuse of RLHF. To start, disinformation campaigns bear an innate limitation on their scale and scope. What makes disinformation effective? Beyond content generation, a successful effort at disinformation is heavily dependent on administration. While automation may free more humans to work on these tasks, propagation of content requires financial and technical infrastructure \cite{CSET1}. Perhaps the best mitigation for misuse of RLHF is to address governance of such infrastructure \cite{CSET1}. Notably, those who control such infrastructure may wield disproportionate power over the direction of RLHF applications.

More broadly, methods to counter AI disinformation will likely be equally useful for RLHF. Cooperation and intelligence sharing between governments and industry parties will be key to developing early warning systems for disinformation campaigns, enabling rapid response, threat information sharing, and cross-platform defense \cite{CSET1}. Since openly released research always carries the potential for misuse, AI researchers must develop more formalized guidelines for guarding against misuse and recommending mitigations. There must be a process by which media outlets can report on disinformation without amplifying its effects. Finally, public resistance to ML-enabled disinformation must be boosted by improving media literacy and increasing the accuracy of public conceptions of AI \cite{CSET1, CSET5, ME6}. At present, ``A good deal of fear and concern about uncontrollable AI is now being displayed in public discourse,’’ which has led to confusion regarding autonomy and the critical role of humans throughout the AI development and deployment pipeline \cite{ME6}. Researchers must be transparent and understandable in how their work is communicated to the public, and media outlets must avoid misleading or over-sensationalized journalism about AI. At the same time, the risk of manipulation can be addressed by increasing digital literacy to enhance personal autonomy and awareness among the general public. 

\subsection{Strengthening Value Alignment}
A core goal of AI research is to produce systems that behave in ways consistent with human values and intentions. There exist varying definitions of AI alignment with human intentions. Askell et al. (2021) define a well-aligned AI to be one that is ``helpful, honest, and harmless,’’ with the understanding that ``these are subtle and ambiguous criteria, and the best AI behavior will involve a compromise between them’’ \cite{ME1}. Current ML approaches tend to suffer from misalignment between the objectives of resulting systems and human values \cite{RLHF9}. Even if we were to observe model behavior fully consistent with human preferences (outer alignment), it is difficult to guarantee true inner alignment\footnote{In the context of AI safety, an \textit{inner alignment} failure refers to any situation in which an AI agent optimizes for goals or objectives different from those we have asked of it \cite{INTERNALALIGNMENT}.} without ulterior motives \cite{RLHF2, INTERNALALIGNMENT}.

RLHF is an important step forward in aligning AI systems with human values as it provides more nuanced guidance than traditional ML and RL, which struggle to capture the full extent of human preference. \cite{GML1}. Specifically, RL algorithms learn the highest-reward path toward a stated objective, sometimes involving actions that lend to economic or physical harm \cite{ME2}. In contrast, there is evidence that RLHF trains models to act in accordance with both explicit (following instructions) and implicit (staying truthful, unbiased, and unhurtful) intentions \cite{GML1}. More broadly, RLHF is an important vantage point for exploring improved systems of value alignment. Insights gained through RLHF are likely transferable to other alignment methods \cite{RLHF7}. Even if RLHF does not completely resolve concerns over inner alignment, the failures it identifies and knowledge it confers to reward and policy modeling are applicable to enhancing safety, reliability, and trustworthiness of AI in social and collaborative situations \cite{RLHF7}.

As AI becomes democratized, how do we construct systems that are sensitive to a diversity of perspectives and value systems and properly aligned in such contexts? Can we design a unified values framework, or should value alignment be restricted to cultural-specific contexts, much like law enforcement differences between nations? Could RLHF make inter- and intra-regional differences in conceptions of morality and ethics more salient? A significant factor in the net impact of RLHF models lies in to whom such models are aligned \cite{GML1}. Challenges exist in designing an alignment process that is fair, unbiased, and transparent while also bearing suitable accountability mechanisms \cite{GML1}. This is relevant in light of unresolved questions over how fundamentally conflicting feedback, values, and preferences should be reconciled, and the fact that there is no consensus across society on any single unified moral theory. Gabriel (2020) suggests pursuing a principle-based approach, whereby models are built to reflect fair principles endorsed by all despite variation in moral beliefs \cite{GABRIEL}. Another path forward concerns training models that are aligned with general principles and preferences, with use of subsequent fine-tuning to condition models to preferences of specific groups. Additional issues are raised by the fact that developer choices can unintentionally impact the behavior of RLHF methods \cite{ME4}. It is perhaps more useful to develop RLHF under assumptions of moral uncertainty, which supposes for any decision that one’s motives are driven by several plausible ethical theories \cite{ME5}. Further consideration must be given to the broad question: should AI agents be able to exhibit the myriad moral and ethical convictions espoused by humans?

\subsection{Mitigating Bias}
With proper deployment, RLHF can reduce bias at multiple levels in the AI production pipeline. Broadly speaking, AI is affected at multiple levels of development by historical bias which affects data generation, representation bias which affects sampling and population studies, measurement bias due to inaccurate data stemming and structural discrimination against groups, aggregation bias due to over-reliance on one-size-fits all models, learning and evaluation bias during model training, and deployment bias due to disparity between intended and observed application \cite{CS1}. Preliminary analysis of RLHF results suggests it can be leveraged to mitigate long-standing effects of historical, representation, and measurement bias by balancing human feedback with representation and expertise across a diverse range of human annotators \cite{morals, ME4}. RLHF is not unsusceptible to bias or misuse, but it leverages human feedback to counter algorithmic bias directly and efficiently in comparison to existing approaches \cite{GML1, morals}. In this light, RLHF is an important tool not only for its potential to transform AI capabilities, but also toward combating systemic inequality issues perpetuated by algorithmic development.

\subsection{Improving Equitable Access and Privacy in AI}
By reducing computational cost, RLHF can open the door to democratization of AI technologies to all levels of society regardless of level of development. In particular, RLHF yields smaller models requiring less compute to achieve state-of-the-art performance \cite{GML1}, which is critical for building practical AI technologies that are deployable across the world and especially to lower-income areas and developing nations. The reduced need for training data can mitigate concerns around data scraping and privacy, security, and surveillance, all of which are issues involved in traditional ML \cite{GML1}. Data collection often disproportionately impacts vulnerable groups in negative ways: data may be misused by technology companies and governments to, for example, track immigrants, and instances of surveillance used to solidify systemic discrimination against subpopulations are well-documented \cite{ME2}. RLHF thus makes it easier to achieve better outcomes without significantly compromising privacy.

\subsection{Bridging Cultures}
RLHF has potential to transform how we reconcile cross-cultural perspectives and approach peaceful dialogue. Cross-cultural feedback is critical to ensuring technology is deployable in contexts beyond domestic production. By soliciting human feedback that encompasses a diversity of viewpoints and cultural norms, RLHF technologies can be culturally aware and usable beyond narrow, culture-specific settings. Even minor cultural awareness can facilitate communication in a number of contexts. A salient example is in education. Mitigating stress associated with feedback interactions in learning is critical to supporting student education \cite{BIAS1}. Yet studies have shown that cross-cultural feedback conversations between teachers and students can compound stress and lead to reduced learning (e.g., via decreased ability to ask questions, ``absorb information, and develop professional and mentoring relationships’’), worsened long-term education outcomes, and increased cognitive load for teachers if approached incorrectly \cite{BIAS1}. This was further exacerbated for interactions between teachers of well-represented identities and students from underrepresented groups \cite{BIAS1}. In this context, RLHF technologies can help overcome such difficulties, whether by moderating conversation or suggesting appropriate ways to approach cross-cultural communication. This benefit extends beyond education into sectors such as customer service and entertainment.

\subsection{Boosting Industries}
By allowing AI agents to learn from human expertise, RLHF can facilitate development of more adaptable AI systems for use in various industries \cite{RLHF8}. Potential applications of RLHF include enhanced resource management, customer service, online education, eldercare, and clinical decision support \cite{EXAMPLES}. Adaptive recommendations could better account for personal and cultural preferences and human intentions; value-aligned technologies could better accommodate individual preferences regarding communication, mobility, and living habits; human-guided diagnostics could improve clarity in decision-making. RLHF can better foster trust with users in order to boost business outcomes across industries and accelerate technology adoption to improve efficiency and economic output.

Concurrently, RLHF can possibly heighten big-tech’s advantage and hasten progress towards dangerous AI capabilities. Notable RLHF advances have been achieved by well-financed research laboratories and big-tech companies such as OpenAI and DeepMind, which can afford to spend enormous amounts of money on creating large datasets for RLHF algorithms. Smaller organizations lack access to such resources \cite{RLHF5}. A related concern regards who should have access to powerful RLHF models produced by organizations. If RLHF models are open-sourced, it may be difficult to check harmful applications and enforce regulation \cite{RLHF5}. Yet restricting access via closed-source models could exclude access to select groups, reducing equity. Likewise concerning is the use of RLHF for weapons development---e.g., better missile systems and more lethal drones. This is a concern for most AI technologies, and global regulatory action must be taken to mitigate possible harm. Lastly, it must be noted that RLHF methods are still susceptible to generic ML vulnerabilities such as adversarial attack, which may affect its ability to enhance industry applications \cite{CSET3}. Awareness of all of these possibilities is critical as RLHF continues to develop.

\subsection{Transforming Work}
RLHF will impact the degree to which different jobs are susceptible to automation. Although many uses of RLHF are still nascent, these developing applications can provide insight into key implications of RLHF for the workforce. With better models that can be more efficiently used, RLHF advances the narrative that RL-based technologies will quickly close the gap between automation and the dexterity and mobility required for low-wage jobs \cite{ME2}. This holds especially for domains in which robotic manipulation and navigation are becoming more dominant. Even so, RLHF is unlikely to lead to full automation of jobs \cite{ME2}. Importantly, RLHF methods can automate tedious or high-risk portions of manual labor \cite{ME2}, especially for tasks which are dangerous or difficult for humans to complete even if they have the correct intuitions \cite{RLHF2}. Humans may guide AI systems in such contexts by providing feedback on how to best complete such tasks \cite{RLHF2}. This can enhance workforce safety and morale and does not fully remove humans from the equation, instead shifting human expertise to different areas of production.

RLHF may further affect the spatial distribution of jobs in the workforce. Resulting automation may move jobs, dependent on factors such as required expertise and closeness to service providers \cite{ME7}. Job relocation in such contexts is not necessarily constrained by national boundaries, exemplified by techniques involving offshoring of automated operations, which, while cost effective, may introduce regulatory challenges, reduce domestic jobs, and impact transparency \cite{ME7}. Future regulations on AI technologies will likely impact the extent to which such impacts are realized.

\section{Further Considerations}
What role should AI play in our daily lives? Critical to answering this question is the related query: ``is AI augmenting human decision, informing it, or supplanting it?’’ \cite{CSET7} RLHF simplifies this evaluation. While most AI applications embody a variation of the \textit{centaur’s dilemma}---the fundamental opposition between human control and optimized AI functionality \cite{CSET7}---RLHF directly plants human feedback as an informative source, leading to greater clarity regarding the locus of human control while simultaneously enhancing functional results. This suggests RLHF is a significant step toward resolving the dilemma, allowing us to reap the full benefits of AI’s capacity and inform rather than undermine human decision-making. Relatedly, many of the positive impacts of RLHF are dependent on the ability to arrive at well-designed human feedback systems. There will inevitably be new ways invented for humans to meaningfully provide feedback to robots and AI agents, as well as new insights as to how human behavior inherently and subtly reveals information signals at any given point \cite{RLHF4}. How will agents extract and make sense of various sources of information? Choose between multiple forms of feedback (for example: comparisons, demonstrations, corrections, improvement, proxy rewards, punishments, credit assignment, linguistic instructions \cite{RLHF4})? Distinguish purposeful from meaningless feedbacks? These considerations will become increasingly important as RLHF advances. Ultimately, the potential for RLHF to positively impact society should not be ignored, and dependence of its benefits on well-designed feedback systems is a further call for investment into RLHF. 

\section{Concluding Remarks}
In this paper, we analyzed the social benefits and harms of RLHF, which is presently one of the foremost and promising AI methods. Specifically, we described how RLHF may net positively impact areas of misinformation, AI value-alignment, bias, equitable access, cross-cultural dialogue, industry, and workforce. This analysis is timely and necessary, as progress on RLHF can impact all levels and sectors of society. Overall, the application of RLHF is important from both safety and capability perspectives. The benefits RLHF projects to provide over the status quo suggests we will see more resources invested in its development. As RLHF raises concerns that echo those of existing AI technologies for governance, industry, safety, ethics, and the future of global power relations, it will be important for all to be aware and intentional in the adoption of RLHF.

\section*{Acknowledgments}
The author would like to recognize and thank Dr. Marion Boulicault for valuable discussion and feedback in the preparation of this paper. This work was funded by the Social and Ethical Responsibilities of Computing at the MIT Schwarzman College of Computing.

\bibliographystyle{unsrt}
\bibliography{biblio}
\nocite{*}



\end{document}